\documentclass[12pt,preprint,showpacs,preprintnumbers]{revtex4}
\usepackage{amsfonts}
\usepackage{float}
\usepackage{amsmath}
\usepackage{amssymb}
\usepackage{hyperref}
\usepackage{color}
\usepackage{graphicx}
\usepackage{amssymb}
\usepackage{amsmath}
\usepackage{graphicx}
\usepackage{dcolumn}
\usepackage{bm}
\usepackage{epsfig}
\usepackage[T1]{fontenc}
\usepackage{ae,aecompl}

\begin{document}
\title{Thermal cascaded lattice Boltzmann method}
\author{ Linlin Fei$^{1}$, K. H. Luo$^{1,2}$\footnote{Corresponding author: K.Luo@ucl.ac.uk}}
\affiliation{$^1$ Center for Combustion Energy, Key laboratory for Thermal Science and Power Engineering of Ministry of Education, Department of Thermal Engineering, Tsinghua University, Beijing 100084, China \\
$^2$ Department of Mechanical Engineering, University College London, Torrington Place, London WC1E 7JE, UK\\
}
\date{\today }

\begin{abstract}
In this paper, a thermal cascaded lattice Boltzmann method (TCLBM) is developed in combination with the double-distribution-function (DDF) approach.
A density distribution function relaxed by the cascaded scheme is employed to solve the flow field, and a total energy distribution function relaxed by the BGK scheme is used to solve temperature field, where two distribution functions are coupled naturally. The forcing terms are incorporated by means of central moments, which is consistent with the previous force scheme
[Premnath  \emph{et al.}, Phys. Rev. E \textbf{80}, 036702 (2009)] but the derivation is more intelligible and the evolution process is simpler. In the method, the viscous heat dissipation and compression work are taken into account, the Prandtl number and specific-heat ratio are adjustable, the external force is considered directly without the Boussinesq assumption, and the low-Mach number compressible flows can also be simulated. The forcing scheme is tested by simulating a steady Taylor-Green flow, and the new TCLBM is then tested by numerical experiments of the thermal Couette flow, low-Mach number shock tube problem, and Rayleigh-B\'{e}nard convection. The simulation results agree well with the analytical solutions and/or results given by previous researchers.
\end{abstract}

\pacs{47.11.-j, 05.20.Dd}
\maketitle

\section{Introduction}
The lattice Boltzmann method (LBM), based on the simplified kinetic models, has gained remarkable success as an alternative method to computational fluid dynamics (CFD) during the past two decades, with applications (but not limited to) to micro flows, flows in porous media, turbulence, magneto-hydrodynamics, reactive flows, and multiphase flows(see e.g.\cite{Chenshiyi_review,Succi_book,Qianyuehong_review} and references therein). Different from solving the discretized Navier-Stokes (N-S) equations in traditional CFD methods, the LBM solves a discrete kinetic equation at the mesoscopic scale, designed to reproduce the N-S equations in the macroscopic limit. The main advantages for LBM over traditional CFD includes \cite{Succi_2008,Liqing_PRO_COMB_2016}: natural incorporation of micro and meso-scale physics, the easiness of programming, the convenience to deal with complex boundary, and high parallel efficiency. Moreover, it has been recently demonstrated \cite{Xuaiguo_2015,Linchuandong_2016} that LBM can be used to develop kinetic models for analyses beyond the traditional hydrodynamic modeling of complex flows.

The basic algorithm realization of LBM is collision-streaming or streaming-collision, although other time and space evolution schemes can also be used. To be specific, at each time step the collision is first locally executed and followed by streaming the post-collision distributions to their neighbors, or just exchanging the above procedure \cite{Guo_book}.
Based on this algorithm, various collision operators can be adopted, such as single-relaxation-time (SRT) operator \cite{Qian_D2Q9}, two-relaxation-time (TRT) operator \cite{I.G_TRT1,I.G_TRT2}, multiple-relaxation-time (MRT) operator \cite{MRT1,MRT2}, entropic operator \cite{karlin_1999a,karlin_1999b,karlin_2000}. Compared with these extensively used operators, cascaded or central moment operator, first proposed by Geier \emph{et al}. \cite{Geier_2006} in 2006, is more recent. The collision in the cascaded Lattice Boltzmann method (CLBM) is performed by relaxing central moments to their local equilibrium values separately, which is different from MRT LBM where the raw moments are relaxed. The Galilean invariance is naturally guaranteed in this central-moments-based LBM. Moreover, as mentioned in \cite{Geier_2006}, central moments can be expressed as polynomials of raw moments of the same order and below by using the binomial theorem, which implies relaxing raw moments (in MRT) effects higher order central moments. This ``cross-talk" is a source of instability and can be removed in CLBM. By choosing the relaxation parameters properly, CLBM can be adopted to simulate very high Reynolds number flows using coarse grids without adopting any turbulence models or entropic stability \cite{Geier_2006}. In 2009, Premnath and Banerjee \cite{Premnath_2009} derived the incorporation of forcing terms in the CLBM by matching the continuous central moments of the force effect with its discrete forms. The forcing terms obtained are thereby Galilean invariant, and the implicitness to apply the trapezoidal rule is removed by introducing a transformed distribution. The second-order accuracy of the above force scheme and some advantages of CLBM over MRT LBM were verified by Ning \emph{et al}. \cite{Premnath_2016}. Recently, Lycett-Brown and Luo \cite{Luokaihong_2014a} extended the CLBM to multiphase flow using the interaction potential method \cite{Shanchen_1993} with EDM force scheme \cite{EDM_force}. Compared with the LBGK method, the proposed model provided significant improvement in reducing spurious velocities, and increasing the stability range for the Reynolds number and liquid to gas density ratio. They further extended the model to three dimensions and made a breakthrough in binary droplet collisions with high Weber number, high Reynolds number and high density ratio simultaneously \cite{Luokaihong_2014b}. It is worth mentioning that Lycett-Brown and Luo's derivation of the CLBM is simpler than that proposed by \emph{et al}. \cite{Geier_2006} and may guide future work in this frame.

Although CLBM has obtained success in high Reynolds number flows and multiphase flows, its applications are so far limited in isothermal fluids. The purpose of the present study is to extend CLBM to thermal flows. Generally, there are three feasible ways to construct thermal LBMs.
The first one, multispeed approach \cite{Alexander_1993,Qianyuehong_1993}, is a straightforward extension of althermal on isothermal LBMs, in which more discrete velocities are adopted to match higher-order moments constraint of the density distribution function for recovering the energy equation. In the second one, a density distribution function is still used to simulate the velocity field, while other methods, such as finite-difference or finite volume \cite{Lallemand_2003a,Lallemand_2003b}, are adopted for the temperature field. Double-distribution-function (DDF)  \cite{He_1998,Guo_2007} approach is the third one, where two different distribution functions are adopted to solve flow and temperature fields, respectively. In DDF-based thermal LBMs, the compression work and heat dissipation can be simply included, and the specific-heat ratio and Prandtl number are adjustable. On the whole, the DDF approach keeps the intrinsic features and simple structure of the standard LBM, and more comparisons and discussions among the three ones can be found in \cite{Guo_2007,Liluo_2012,Karlin_2013}. In history, the first DDF thermal model was proposed by He \emph{et al}. \cite{He_1998} by using an internal-energy-distribution-function-based DDF approach. Guo \emph{et al}. \cite{Guo_2007} then presented another DDF thermal model using total energy distribution function to solve the energy equation, which is simpler than He and co-workers' model. In Guo and co-authers' model, the local temperature in equilibrium density and energy distribution functions is replaced by the reference temperature, thus it is a decoupling model and is limited to Boussinesq flows. In 2012, Li \emph{et al}. \cite{Liluo_2012} developed a coupling DDF thermal model which can simulate more general thermal flows, and the model was extended to three-dimension by Feng \emph{et al}. \cite{Feng_2015}  recently. Inspired by these works, we try to construct a thermal cascaded lattice Boltzmann method (TCLBM) in the present work based on the DDF approach. In the TCLBM, a density distribution function is relaxed using the cascaded scheme, a total energy distribution function is relaxed using the SRT scheme, and the forcing terms are incorporated by means of central moments.

The rest of the paper is structured as follows: Section \ref{sec.2} briefly introduces the cascaded LBM. Section \ref{sec3} presents an incorporation method for force effect that is different from the original one \cite{Premnath_2009}. In Sec. \ref{sec4}, we extend the athermal CLBM to TCLBM in detail. Numerical experiments are carried out for several benchmark problems to validate the proposed model in Sec. \ref{sec5}. Finally, conclusions of this work are made in Sec. \ref{sec.6}.

\section{Cascaded LBM}\label{sec.2}

Without losing the generality, the D2Q9 lattice \cite{Qian_D2Q9} is adopted here, and the discrete velocities are defined as $\mathbf{e}_{0}=(0,0)$, 
$\mathbf{e}_{a}=(\cos[(a-1)\pi/2],\sin[(a-1)\pi/2])c$, for $a=$1\textendash4, and $\mathbf{e}_{a}=(\sqrt{2}\cos[(a-9/2)\pi/2],\sqrt{2}\sin[(a-9/2)\pi/2])c$ 
 for $a=$5\textendash8. In LBM, $ c=\Delta{x}/\Delta {t}=1 $, $ c_{s}=1/\sqrt{3} $, here $ \Delta{x} $ and $ \Delta {t} $ are the lattice spacing and time step.
For the derivation of CLBM, we follow Lycett-Brown and Luo \cite{Luokaihong_2014a} and begin with the velocity moments of the discrete distribution function (DF) $ f_a $, and then $ f_a $ and  $ f_a^{eq} $ can be formulated as functions of the corresponding moments and equilibrium moments.

The raw moments are defined as
\begin{equation}\label{e1}
\rho M_{pq}=\sum\limits_{a}f_{a}\mathit{e}_{ax}^{p}\mathit{e}_{ay}^{q}.
\end{equation}
in this notation, the zero-order moment $M_{00}=1$ , and first-order moments $M_{10}=u_x$ and $M_{01}=u_y$ are conserved, corresponding to mass and momentum conservations, respectively. To get the formulations of  $ f_a $ , another six independent moments are needed, including three second-order moments ($M_{11}$, $M_{02}$ and $M_{20}$), two third-order moments ($M_{21}$, $M_{12}$, noting that $M_{03}$ and $M_{30}$ are not independent of the first-order ones owing to the lack of symmetry in D2Q9 lattice), and the fourth-order moment $M_{22}$. Recombining the second-order moments, the trace of the pressure tensor, the normal stress difference and the off diagonal element of the pressure tensor are given by
\begin{equation}\label{e2}
E=M_{20}+M_{02},  ~~N=M_{20}-M_{02},  ~~\varPi=M_{11}.
\end{equation}
According to the definition above, we get the raw moment representation of populations:
\begin{subequations}\label{e3}
	\begin{equation}\label{e3a}
	{f_0} = \rho \left[ {{M_{00}} - E + {M_{22}}} \right],	
	\end{equation}
	\begin{equation}
	{f_1} = \frac{1}{2}\rho \left[ {{M_{10}} + \frac{1}{2}(E + N) - {M_{12}} - {M_{22}}} \right],
	\end{equation}
	\begin{equation}
	{f_2} = \frac{1}{2}\rho \left[ {{M_{01}} + \frac{1}{2}(E - N) - {M_{21}} - {M_{22}}} \right],
	\end{equation}
	\begin{equation}
	{f_3} = \frac{1}{2}\rho \left[ { - {M_{10}} + \frac{1}{2}(E + N) + {M_{12}} - {M_{22}}} \right],
	\end{equation}
	\begin{equation}
	{f_4} = \frac{1}{2}\rho \left[ { - {M_{01}} + \frac{1}{2}(E - N) + {M_{21}} - {M_{22}}} \right],
	\end{equation}
	\begin{equation}
	{f_5} = \frac{1}{4}\rho \left[ {{\varPi} + {M_{21}} + {M_{12}} + {M_{22}}} \right],
	\end{equation}
	\begin{equation}
	{f_6} = \frac{1}{4}\rho \left[ { - {\varPi} + {M_{21}} - {M_{12}} + {M_{22}}} \right],
	\end{equation}
	\begin{equation}
	{f_7} = \frac{1}{4}\rho \left[ {{\varPi} - {M_{21}} - {M_{12}} + {M_{22}}} \right],
	\end{equation}
	\begin{equation}
	{f_8} = \frac{1}{4}\rho \left[ { - {\varPi} - {M_{21}} + {M_{12}} + {M_{22}}} \right].
	\end{equation}
\end{subequations}
It should be noted that other variables can also be expressed using their moments in this form similarly.

Central moments are defined in a reference frame shifted by the local velocity,
\begin{equation}\label{e4}
\rho {\tilde M_{pq}} = \sum\limits_a {{f_a}} {({e_{ax}} - {u_x})^p}{({e_{ay}} - {u_y})^q}.
\end{equation}
The transformation between the row moments and central moments can be expressed using the binomial theorem as given by Lycett-Brown and Luo \cite{Luokaihong_2014a}. To construct a CLBM, we follow the assumption adopted in \cite{Premnath_2009}, by setting the discrete equilibrium central moments equal to the corresponding continuous values,
\begin{equation}\label{e5}
\rho {\tilde M^{eq}}_{pq} = \int_{ - \infty }^\infty  {\int_{ - \infty }^\infty  {{f^{eq}}{{({\xi _x} - {u_x})}^p}} } {({\xi _y} - {u_y})^q}d{\xi _x}{\xi _y}
\end{equation}
where ${{f^{eq}}}$ is the local Maxwell-Boltzmann distribution for athermal fluid at temperature $ T_0 $ in continuous particle velocity space $ ({\xi _x},{\xi _y})$,
\begin{equation}\label{e6}
{f^{eq}} = \frac{\rho }{{2\pi R{T_0}}}\exp \left[ { - \frac{{{{(\xi - \mathbf{u})}^2}}}{{2\pi R{T_0}}}} \right].
\end{equation}
Substituting Eq. (\ref{e6}) into Eq. (\ref{e5}), we can calculate the second order and above central moments, and write them using the combination as done in raw moments:
\begin{equation}\label{e7}
\tilde \varPi^{eq} = {\tilde N^{eq}} = \tilde M_{21}^{eq} = \tilde M_{12}^{eq} = 0,~~{\tilde E^{eq}} = 2R{T_0},~~{\tilde M^{eq}}_{22} = {(R{T_0})^2}.
\end{equation}

The implementation of CLBM is also composed of collision step and streaming step. For the collision step, central moments are relaxed to their equilibrium values, separately:
\begin{subequations}\label{e8}
	\begin{equation}
	{\tilde \varPi ^ * } = {w_1}{\tilde \varPi ^{eq}} + (1 - {w_1})\tilde \varPi,
	\end{equation}
	\begin{equation}
	{\tilde N^ * } = {w_1}{\tilde N^{eq}} + (1 - {w_1})\tilde N,
	\end{equation}
	\begin{equation}
	{\tilde E^ * } = {w_2}{\tilde E^{eq}} + (1 - {w_2})\tilde E,
	\end{equation}
	\begin{equation}
	\tilde M_{21}^ * = {w_3}\tilde M_{21}^{eq}{\rm{ + }}(1 - {w_3})\tilde M_{21},
	\end{equation}
	\begin{equation}
	\tilde M_{12}^*  = {w_3}\tilde M_{12}^{eq}{\rm{ + }}(1 - {w_3})\tilde M_{12},
	\end{equation}
	\begin{equation}
	\tilde M_{22}^ *  = {w_4}\tilde M_{22}^{eq}{\rm{ + }}(1 - {w_4})\tilde M_{22},
	\end{equation}		
\end{subequations}
where $ w_1 $ and $ w_2 $ are dependent on the shear and bulk viscosities,  respectively($ \nu {\rm{ = }}R{T_0}(1/{w_1} - 0.5)
 $, and $ {\nu _B}{\rm{ = }}R{T_0}(1/{w_2} - 0.5)
  $), and the parameters for the third- and fourth-central moments ($ w_3 $ and $ w_4 $) are freely tunable to control the stability. The post-collision raw moments can then be recovered according to  the binomial theorem,
  \begin{subequations}\label{e9}
  	\begin{equation}
  	{\varPi ^ * } = {\tilde \varPi ^ * } + {u_x}{u_y},
  	\end{equation}
  	\begin{equation}
    {N^ * } = {\tilde N^ * } + u_x^2 - u_y^2,
  	\end{equation}
  	\begin{equation}
	 {E^ * } = {\tilde E^ * } + u_x^2 + u_y^2,
  	\end{equation}
  	\begin{equation}
  	M_{21}^ *  = \tilde M_{21}^ *  + 2{u_x}{\varPi ^ * } + \frac{1}{2}{u_y}({E^ * } + {N^ * }) - 2u_x^2{u_y},	
  	\end{equation}
  	\begin{equation}
  	M_{12}^ *  = \tilde M_{12}^ *  + 2{u_y}{\varPi ^ * } + \frac{1}{2}{u_x}({E^ * } - {N^ * }) - 2u_y^2{u_x},
  	\end{equation}
  	\begin{equation}
  	M_{22}^ *  = \tilde M_{22}^ *  + 2{u_x}M_{12}^ *  + 2{u_y}M_{21}^ *  - \frac{1}{2}{\mathbf{u}^2}{E^ * } + \frac{1}{2}(u_x^2 - u_y^2){N^ * } - 4{u_x}{u_y}{\varPi ^ * } + 3u_x^2u_y^2.
  	\end{equation}
  \end{subequations}
Then we get the post-collision distribution using Eq. (\ref{e3}), and the streaming step is as usual,
\begin{equation}
{f_a}(\mathbf{x} + {\mathbf{e}_a},t + 1) = f_a^ * (\mathbf{x},t).
\end{equation}

\section{INCORPORATING FORCING TERMS INTO CASCADED LBM}\label{sec3}
To include the force effect on the flow field, we define $ f_a $ changes due to this force field by a source term $ {S_a}$. To match the overall accuracy in LBM, one way to add the source term in CLBM is to employ the second-order trapezoidal rule along the characteristic line,
\begin{equation}\label{e11}
{f_a}(\mathbf{x} + {\mathbf{e}_a},t + 1) = f_a^ * (\mathbf{x},t) + \frac{1}{2}\left[ {{S_{a(\mathbf{x},t)}} + {S_{a(\mathbf{x} + {\mathbf{e}_a},t + 1)}}} \right].
\end{equation}
To remove the implicitness in Eq. (\ref{e11}), the transformation method in \cite{Premnath_2009} is adopted,
\begin{equation}\label{e12}
{\bar f_a}(\mathbf{x} + {\mathbf{e}_a},t + 1) = \bar f_a^ * (\mathbf{x},t) + {S_{a(\mathbf{x},t)}},~~{\bar f_a} = {f_a}{\rm{ - }}\frac{1}{2}{S_a}.
\end{equation}
He \emph{et al}. \cite{He_1998} proposed that the presence of the force field $ \mathbf{F} = (F_x,F_y)$ changes the continuous distribution function as follows:
\begin{equation}\label{e13}
\Delta {f^\mathbf{F}} = \frac{\mathbf{F}}{\rho } \cdot \frac{{(\xi  - \mathbf{u})}}{{R{T_0}}}{f^{eq}}.
\end{equation}
We then follow the assumption in \cite{Premnath_2009} that the discrete central moments of $ {S_a}$ are equal to the continuous central moments of $ \Delta {f^\mathbf{F}}$:
\begin{equation}
\rho \tilde M_{pq}^s = \sum\limits_a {{S_a}} {({e_{ax}} - {u_x})^p}{({e_{ay}} - {u_y})^q} = \int_{{\rm{ - }}\infty }^\infty  {\int_{{\rm{ - }}\infty }^\infty  {\Delta {f^\mathbf{F}}{{({\xi _x} - {u_x})}^p}{{({\xi _y} - {u_y})}^q}d{\xi _x}{\xi _y}} }. 
\end{equation}
Substituting Eq. (\ref{e13}) into the integral, we get the nonzero central moments of $ {S_a}$,
\begin{subequations}\label{e15}
	\begin{equation}
	\tilde M_{10}^s = {a_x},
	\end{equation}
	\begin{equation}
	\tilde M_{01}^s = {a_y},	
	\end{equation}
	\begin{equation}
	\tilde M_{21}^s = R{T_0}{a_y},
	\end{equation}
	\begin{equation}
	\tilde M_{12}^s = R{T_0}{a_x}.
	\end{equation}
\end{subequations}
where $ a_x $ and $ a_y $ are horizontal and vertical components of the acceleration. As mentioned by Premnath and Banerjee \cite{Premnath_2009}, the third-order moments have no effects on the recovered Navier-Stokes equations, thus we will remove them henceforward for convenience. Using the binomial theorem once again, we yield analytical raw moments of $ {S_a}$,
\begin{subequations}
	\begin{equation}
	M_{00}^s = 0,
	\end{equation}
	\begin{equation}
	M_{10}^s = {a_x},
	\end{equation}
	\begin{equation}
	M_{01}^s = {a_y},
	\end{equation}
	\begin{equation}
	{E^s} = 2({a_x}{u_x} + {a_y}{u_y}),
	\end{equation}
	\begin{equation}
	{N^s} = 2({a_x}{u_x} - {a_y}{u_y}),	
	\end{equation}
	\begin{equation}
	{\varPi ^s} = {a_x}{u_y} + {a_y}{u_x},
	\end{equation}
	\begin{equation}
    M_{21}^s = {a_y}u_x^2 + 2{a_x}{u_x}{u_y},
	\end{equation}
	\begin{equation}
	M_{12}^s = {a_x}u_y^2 + 2{a_y}{u_x}{u_y},
	\end{equation}
	\begin{equation}
	M_{22}^s = 2{a_x}{u_x}u_y^2 + 2{a_y}{u_y}u_x^2.
	\end{equation}
\end{subequations}
Thus, the analytical expressions of $ {S_a}$ can be written in the same form as Eq. (\ref{e3}).

From the definition in Eq. (\ref{e12}), the conserved raw moments of the transformed discrete distribution $ {\bar f_a} $ are  $ {\bar M_{00}} = 1
 $, $ {\bar M_{10}} = {u_x} - 0.5{a_x}$, and $ {\bar M_{01}} = {u_y} - 0.5{a_y}$,  respectively. The corresponding non-conserved raw and central moments can then be obtained straightforwardly,
\begin{equation}
{\bar{M}_{pq}} = {M_{pq}} - {\kern 1pt} \frac{1}{2}M_{pq}^s,~~{\tilde{\bar M}_{pq}} = {\tilde M_{pq}},~~(p + q >  = 2).
\end{equation}
With Eqs. (\ref{e7}) and (\ref{e15}), the non-conserved equilibrium central moments will remain the same as the ones before transformed, thus the collision step for the central moments will not be affected in Eq. (\ref{e8}). According to the relationship between raw moments mentioned above, the post-collision raw moments are slightly different from Eq. (\ref{e9}),
\begin{subequations}\label{e18}
	\begin{equation}
	\bar \varPi  = {\tilde{\bar \varPi}^ * } + {u_x}{u_y}{\rm{ - }}\frac{1}{2}({a_x}{u_y} + {a_y}{u_x}),
	\end{equation}
	\begin{equation}
{\bar N^ * } = {\tilde{\bar N}^ *  } + u_x^2 - u_y^2{\rm{ - }}({a_x}{u_x} - {a_y}{u_y}),
	\end{equation}
	\begin{equation}
{\bar E^ * } = {\tilde{\bar E}^ *  } + u_x^2 + u_y^2 - ({a_x}{u_x} + {a_y}{u_y}),
	\end{equation}
	\begin{equation}
	\bar M_{21}^ *  = {\tilde{\bar M}}_{21}^*  + 2{u_x}{\bar \Pi ^ * } + \frac{1}{2}{u_y}({\bar E^ * } + {\bar N^ * }) - 2u_x^2{u_y} + \frac{1}{2}{a_y}u_x^2 + {a_x}{u_x}{u_y}	\label{key},
	\end{equation}
	\begin{equation}
	\bar M_{12}^ *  = {\tilde{\bar M}}_{12}^*    + 2{u_y}\bar\varPi^ *  + \frac{1}{2}{u_x}({\bar E^ * } - {\bar N^ * }) - 2u_y^2{u_x} + \frac{1}{2}{a_x}u_y^2 + {a_y}{u_x}{u_y},
	\end{equation}
	\begin{equation}
	\bar M_{22}^ *  = {\tilde{\bar M}}_{22}^*   + 2{u_x}\bar M_{12}^ *  + 2{u_y}\bar M_{21}^ *  - \frac{1}{2}{\mathbf{u}^2}{\bar E^ * } + \frac{1}{2}(u_x^2 - u_y^2){\bar N^ * } - 4{u_x}{u_y}{\bar\varPi^ * } + 3u_x^2u_y^2{\rm{ - }}{a_x}{u_x}u_y^2 - {a_y}{u_y}u_x^2.
	\end{equation}
\end{subequations}
After substituting Eq. (\ref{e18}) into Eq. (\ref{e3}) together with the conserved ones ($ {\bar M_{00}}
 $, $ {\bar M_{10}}
  $ and $ {\bar M_{01}}
   $)  to get $ \bar f_a^ * 
    $ , the streaming step is then given as:
\begin{equation}
{\bar f_a}(\mathbf{x} + {\mathbf{e}_a},t + 1) = \bar f_a^ * (\mathbf{x} ,t) + {S_{a(\mathbf{x} ,t)}}.
\end{equation}
The hydrodynamic variables are then obtained as:
\begin{equation}\label{e20}
\rho {\rm{  = }}\sum\limits_a {{{\bar f}_a}},~~\rho\mathbf{u} = \sum\limits_a {{{\bar f}_a}}{\mathbf{e}_a} + 0.5\mathbf{F}.
\end{equation}
Now, the force terms are incorporated into the CLBM by means of central moments. It should be noted that the method mentioned above is consistent with the method in \cite{Premnath_2009}, while the derivation is more intelligible and the evolution process is simpler.

\section{COUPLING DDF CASCADED LBM FOR THERMAL FLOW}\label{sec4}
For thermal flow, the temperature $ T $ is now a function of space and time, not a constant value $ T_0 $. The equilibrium distribution function $ f^{eq} $ is given by,
\begin{equation}
{f^{eq}} = \frac{\rho }{{2\pi RT}}\exp \left[ { - \frac{{{{(\xi  - \mathbf{u})}^2}}}{{2\pi R{T_{}}}}} \right],
\end{equation}
and then the reference temperature $ T_0 $ in Eq. (\ref{e7}) should be replaced by local temperature $ T $. Inspired by the total-energy-based DDF models \cite{Guo_2007,Luokaihong_2014a}, we adopt a density distribution function relaxed by the cascaded scheme to solve the flow field, together with a total energy distribution function using the BGK relaxation scheme to simulate the temperature field, and the ideal gas equation of state (EOS, $ p=\rho RT $  ) is used as a bridge to couple them together. The discrete total energy distribution function  has the kinetic equation \cite{Guo_2007},
\begin{equation}\label{e22}
{\partial _t}{h_a} + {\mathbf{e}_a}\cdot\nabla {h_a} =  -\frac{1}{{{\tau _h}}}\left( {{h_a} - h_a^{eq}} \right) + \frac{{{Z_a}}}{{{\tau _{hf}}}}({f_a} - f_a^{eq}), ~~~~{Z_a} = {\mathbf{e}_a} \cdot \mathbf{u} - {\mathbf{u}^2}/2.
\end{equation}

To recover the compressible N-S equations, discrete raw moments of  $ f_a^{eq} $ should be consistent with the continuous raw moments of $ f^{eq} $ from the zeroth- to third-order . As mentioned in Sec. \ref{sec.2}, two of the third-order raw moments are not independent due to the lack of symmetry in D2Q9 lattice,
\begin{subequations}
	\begin{equation}
	\sum\limits_a {f_a^{eq}e_{ax}^3}  = \sum\limits_a {f_a^{eq}e_{ax}}  = \rho {u_x},	
	\end{equation}
	\begin{equation}
\sum\limits_a {f_a^{eq}e_{ax}^3}  = \sum\limits_a {f_a^{eq}e_{ax}}  = \rho {u_y}.
	\end{equation}
\end{subequations}
Combining them with $ M_{21}^{eq} = {u_y}RT + u_x^2{u_y}
 $  and $ M_{12}^{eq} = {u_x}RT + u_y^2{u_x}
  $ , we have,
\begin{equation}
\sum\limits_a {f_a^{eq}{e_{ai}}} {e_{aj}}{e_{ak}} = \rho RT({u_k}{\delta _{ij}} + {u_j}{\delta _{ik}} + {u_i}{\delta _{kj}}) + \rho {u_i}{u_j}{u_k} + \rho {u_l}(1 - \theta  - u_l^2){\delta _{ijkl}}.
\end{equation}
The last term at the RHS is a deviation from the continuous moments for $ f^{eq} $, where $\theta  = T/{T_0}$ , and $ {\delta _{ijkl}} = 1$  when $ i = j = k = l
 $, else $ {\delta _{ijkl}} = 0$. This means that the diagonal elements ($ {\delta _{ijkl}} = 1$) for the-third order velocity moments deviate from the needed relationship. As pointed out by Prasianakis and Karlin \cite{Karlin_2007}, the deviation can be removed only by adding a correction term $ {C_a}
$ into the evolution equation for standard lattices. According to Li and co-workers' work \cite{Liluo_2012}, the raw moments for $ {C_a}$ should satisfy,
\begin{subequations}
	\begin{eqnarray}
	M_{00}^c = M_{10}^c = M_{01}^c = M_{11}^c = 0,
	\end{eqnarray}
	\begin{equation}
{E^c} = {\partial _x}\left[ {{u_x}(1 - \theta )} \right] + {\partial _y}\left[ {{u_y}(1 - \theta )} \right],
	\end{equation}
	\begin{equation}
	{N^c} = {\partial _x}\left[ {{u_x}(1 - \theta )} \right] - {\partial _y}\left[ {{u_y}(1 - \theta )} \right].	
	\end{equation}
	The other moments can be chosen as
	\begin{equation}
	M_{21}^c = {u_y}{\partial _x}\left[ {{u_x}(1 - \theta )} \right],
	\end{equation}
	\begin{equation}
	M_{12}^c = {u_x}{\partial _y}\left[ {{u_y}(1 - \theta )} \right],
	\end{equation}
	\begin{equation}
	M_{22}^c = 0.
	\end{equation}
\end{subequations}
In the above, all the third-order velocity terms have been neglected because of the low Mach number limit. Then all the central moments for $ {C_a}$ are zero except:
\begin{subequations}
	\begin{equation}
	{\tilde E^c} = {\partial _x}\left[ {{u_x}(1 - \theta )} \right] + {\partial _y}\left[ {{u_y}(1 - \theta )} \right],
	\end{equation}
	\begin{equation}
	{\tilde N^c} = {\partial _x}\left[ {{u_x}(1 - \theta )} \right] - {\partial _y}\left[ {{u_y}(1 - \theta )} \right].
	\end{equation}
\end{subequations}
In the simulation, the derivative terms can be evaluated using second-order central difference. Then the analytical expressions of $ {C_a}$ can be written in the same form as Eq. (\ref{e3}).

To remove the implicitness for adding correction terms using a second-order trapezoidal rule, the evolution equation can be written analogously:
\begin{equation}\label{e27}
\bar{f}_a(\mathbf{x} + \mathbf{e}_a,t + 1) = \bar{f}_a^ *(\mathbf{x},t) + {S_{a(\mathbf{x},t)}} + {C_{a(\mathbf{x},t)}}, ~~{\bar f_a} = {f_a}{\rm{ - }}\frac{1}{2}{S_a} - \frac{1}{2}{C_a}.
\end{equation}
Due to the non-zero second-order central moments ($ {\tilde E^c}
 $ and $ {\tilde N^c}
  $ ) for $ {C_a}$, the equilibrium central moments for the transformed distribution $ {\bar f_a}$ should be:
\begin{subequations}
	\begin{equation}
	\tilde{\bar\varPi}^{eq} = \tilde {\bar M}_{21}^{eq} = \tilde {\bar M}_{12}^{eq} = 0,
	\end{equation}
	\begin{equation}
	{\tilde {\bar N}^{eq}} = \frac{1}{2}{\partial _y}\left[ {{u_y}(1 - \theta )} \right] - \frac{1}{2}{\partial _x}\left[ {{u_x}(1 - \theta )} \right],	
	\end{equation}
	\begin{equation}
	{\tilde {\bar E}^{eq}} = 2RT  - \frac{1}{2}{\partial _x}\left[ {{u_x}(1 - \theta )} \right] - \frac{1}{2}{\partial _y}\left[ {{u_y}(1 - \theta )} \right],
	\end{equation}
	\begin{equation}
	{\tilde {\bar M}_{22}^{eq}} = {(RT)^2}.
	\end{equation}
\end{subequations}
The cascaded relaxation for central moments is in the same form as Eq. (\ref{e8}), while the dynamic viscosity $ \mu $ and bulk viscosity $ {\mu _B}$ are:
	\begin{equation}\label{e29}
	\mu  = p(1/{w_1} - 0.5),~~{\mu _B} = p(1/{w_2} - 0.5).	
	\end{equation}
Because the conserved raw moments for  $ {C_a}$  are zero, the calculation for post-collision raw moments is the same as Eq. (\ref{e18}). Then $ \bar{f}_a^ * $  can be obtained using Eq. (\ref{e3}) once again. After the streaming step Eq. (\ref{e27}), hydrodynamic variables are then obtained using Eq. (\ref{e20}).

To recover the total energy equation, the velocity moments for $ h_a^{eq}
 $  should satisfy:
\begin{subequations}
	\begin{equation}
	M_{00}^{heq} = E,	
	\end{equation}
	\begin{equation}
	M_{10}^{heq} = (E + RT){u_x},
	\end{equation}
	\begin{equation}
	M_{01}^{heq} = (E + RT){u_y},
	\end{equation}
	\begin{equation}
	M_{20}^{heq} = (E + 2RT)u_x^2 + RT\left( {E + RT} \right),
	\end{equation}
	\begin{equation}
	M_{02}^{heq} = (E + 2RT)u_y^2 + RT\left( {E + RT} \right),
	\end{equation}
	\begin{equation}
	M_{11}^{heq} = (E + 2RT){u_x}{u_y}.	
	\end{equation}
\end{subequations}
where $ E = (bRT + {\mathbf{u}^{\rm{2}}})/{\rm{2}} $ is the total energy, in which the gas has $ b $ degrees of freedom. Besides, we can set higher raw moments to be zero, then $ h_a^{eq}$  can be given by Eq. (\ref{e3}),
\begin{equation}
h_a^{eq} = \left\{ \begin{array}{l r}
\rho \left[ {E - (E + 2RT){\mathbf{u}^2} - 2RT\left( {E + RT} \right)} \right], & a = 0 \\ 
\frac{1}{2}\rho \left[ {(E + RT)\mathbf{e}_{a} \cdot \mathbf{u} + (E + 2RT)(e_{ax}^2u_x^2+ e_{ay}^2u_y^2){\rm{ + }}RT\left( {E + RT} \right)} \right], &a = 1,...,4 \\
\frac{1}{4}\rho \left[(E + 2RT){e_{ax}}{e_{ay}{u_x}{u_y}} \right], & a = 5,...8 \\ 
\end{array} \right.
\end{equation}
In the same manner, we use a transformed total energy distribution function:
\begin{equation}
{\bar h_a} = {h_a} - \frac{1}{2}{K_a}, ~~{K_a} = \frac{{{Z_a}}}{{{\tau _{hf}}}}(\bar f - f_a^{eq} + \frac{1}{2}{S_a} + \frac{1}{2}{C_a}).
\end{equation}
Then the time-discrete form of Eq. (\ref{e22}) is,
\begin{equation}
{\bar h_a}(\mathbf{x} + {\mathbf{e}_a},t + 1) - {\bar h_a}(\mathbf{x},t) =  - {w_h}\left[ {{{\bar h}_a}(\mathbf{x},t) - h_a^{eq}(\mathbf{x},t)} \right] + (1 - 0.5{w_h}){K_{a(\mathbf{x} ,t)}}, 
\end{equation}
where the relaxation parameters are related to the thermal conductivity $ \lambda $ and Pr number \cite{Liluo_2012},
	\begin{equation}
	\frac{1}{{{w_h}}} = \frac{{2\lambda }}{{pR(b + 2)}} + 0.5,
	~~{\tau _{hf}} = (\mu /p + 0.5)/(\Pr  - 1).
	\end{equation}
The macroscopic temperature is obtained by,
\begin{equation}
T = \frac{2}{{bR}}\left( {\sum\limits_a {{{\bar h}_a}/\rho  - \frac{1}{2}{\mathbf{u}^2}} } \right).
\end{equation}

\section{NUMERICAL EXPERIMENTS}\label{sec5}
In this section, a series of numerical experiments are conducted to verify the developed model. First, a steady Taylor-Green flow is simulated to check the force implementation. Besides, three thermal tests, including thermal Couette flow, low-Mach shock tube problem, and Rayleigh-B\'{e}nard convection, are simulated. To be general, the correction terms in Eq. (\ref{e27}) are considered in all the thermal cases.
\subsection{Steady Taylor-Green flow}
The incompressible N-S equations are as follows,
\begin{subequations}
	\begin{equation}
	\nabla  \cdot {\mathbf{u}} = 0,
	\end{equation}
	\begin{equation}
	{\partial _t}\mathbf{u} + \nabla  \cdot (\mathbf{uu}) =  - \nabla p + \mathbf{F} + \nu \Delta \mathbf{u}.
	\end{equation}
\end{subequations}
In two dimensions, if the force $ \mathbf{F} = ({F_x},{F_y})
 $ is in the form,
 \begin{subequations}
 	\begin{equation}
 	{F_x} = 2\nu \sin (x)\sin (y),	
 	\end{equation}
 	\begin{equation}
 	{F_y} = 2\nu \cos (x)\cos (y).	
 	\end{equation}
 \end{subequations}
Then the solution is in the same form with the Taylor-Green flow but without the time dependency,
\begin{subequations}
	\begin{equation}
	{u_{ax}}(x,y) = {u_0}\sin (x)\sin (y),	
	\end{equation}
	\begin{equation}
	{u_{ay}}(x,y) = {u_0}\cos (x)\cos (y),
	\end{equation}
	\begin{equation}
	p(x,y) = {p_0} + 0.25u_0^2(\cos (2x) - \cos (2y)).	
	\end{equation}
\end{subequations}

In the simulation, the computational domain of $ 0 \le x,y \le 2\pi 
 $ is covered by $ {N_x} \times {N_y}
  $ grid nodes, which means $ \Delta x = 2\pi /{N_x}
   $. The definition of Reynords number is $ {\mathop{\rm Re}\nolimits}  = {u_0}\pi /v = {u_0}{N_x}/(2{v_m})
    $, where the ``real'' kinetic viscosity $ v
     $ is set to be 0.002 in all the cases and the ``model'' kinetic viscosity $ {v_m} $ is related to the relaxation parameter $ w_1 $. Zero velocities and constant density 1.0 are given at every nodes for the initial conditions, while periodic boundary conditions are applied along all the boundaries. To eliminate the artificial compressibility, $ u_0 $ should be set very small.
     
\begin{figure*}[!ht]
 	\center {
 		{\epsfig{file=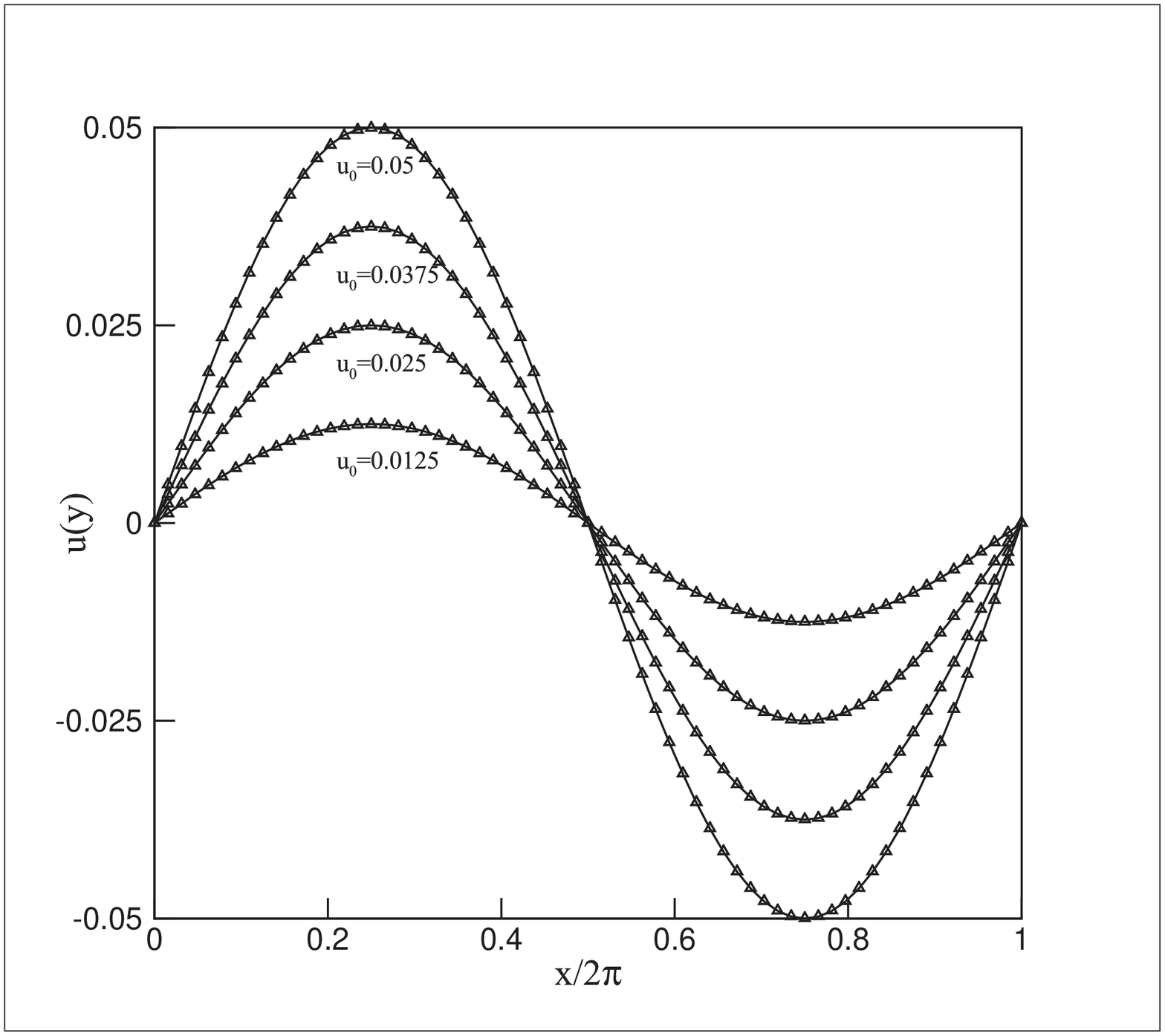,bbllx=20pt,bblly=75pt,bburx=580pt,bbury=580pt,width=0.5\textwidth,clip=}}	
 	}
 	\caption{Comparison of the horizontal velocity profiles at $ y = \pi 
 		$ simulated by the CLBM using the present force scheme (symbols) with the analytical solutions (lines) in different values of $ u_0 $.}
 	\label{FIG1}
 \end{figure*}

First we use $ {N_x}{\rm{ = 64}}
 $ to simulate four cases with $ u_0 $=0.0125, 0.025, 0.0375, and 0.05. In all the cases, $ {w_1} = {w_2} = {w_3} = {w_4} $ is used to recover the BGK collision operator.The residual error $ {E_R} < {10^{ - 8}} $ is used as the convergent criterion, and the realtive error  $ {E_2} $ is calculated for the following analysis,
 \begin{equation}
 {E_R} = \sqrt {\frac{{\sum {{{({\mathbf{u}_{(t + 1000\delta t)}} - {\mathbf{u}_t})}^2}} }}{{\sum {\mathbf{u}_{(t + 1000\delta t)}^2} }}},~~ {E_2} = \sqrt {\frac{{\sum {{{(\mathbf{u} - {\mathbf{u}_a})}^2}} }}{{\sum {{\mathbf{u}_a}^2} }}}.
 \end{equation}
 As shown in Fig. \ref{FIG1}, the simulation results of the horizontal velocity profiles using the CLBM with the present force scheme are in good agreement with the analytical solutions at different velocity amplitudes (Reynolds numbers).
 
 To check the accuary of the present force scheme, the force scheme proposed by Guo \emph{et al.} \cite{Guoforce} is used to make comparisons. As it is illustrated in TABLE \ref{TAB1},  $ E_2 $ of Guo's force scheme with the original EDF (Guo1) has slight deviation from the present scheme when $ u_0=0.0125 $, and with the increasing of $ u_0 $, the deviation becomes more and more severe. It should be noted that the third- and fourth-order velocity terms are included in the EDF of the cascaded LBM, while they are removed in the ``standard" EDF. For fairness, $ E_2 $ of Guo's force scheme with the EDF in the CLBM (Guo2) is also listed in TABLE \ref{TAB1}. As it is shown, the relative errors for Guo2 and the present scheme remains almost constant when $ u_0 $ is increased, and the difference between them is negligibly samll. It can be concluded that the accuracy of Guo's scheme is the same as the present scheme when the same EDF is used for isothermal flows.

\begin{table}[htbp]
	\caption{Comparison of $ {E_2} $ between Guo's force scheme and the present scheme}
	\label{TAB1}
		\begin{tabular}{ c   c   c   c   c}
		\toprule
	$ u_0 $	& 0.0125 & 0.025  & 0.0375 &0.05 \\ \hline
	Guo1	& $ 1.607\times10^{-3} $ & $ 1.628\times10^{-3} $  & $ 1.702\times10^{-3} $ & $ 1.909\times10^{-3} $\\  \hline
	Present	& $ 1.587\times10^{-3} $ & $ 1.554\times10^{-3} $  & $ 1.520\times10^{-3} $ & $ 1.520\times10^{-3} $\\ 
	\hline
	Guo2	& $ 1.581\times10^{-3} $ & $ 1.545\times10^{-3} $  & $ 1.510\times10^{-3} $ & $ 1.510\times10^{-3} $\\
		\toprule	
	\end{tabular}
\end{table}
\begin{figure*}[!ht]
	\center {
		{\epsfig{file=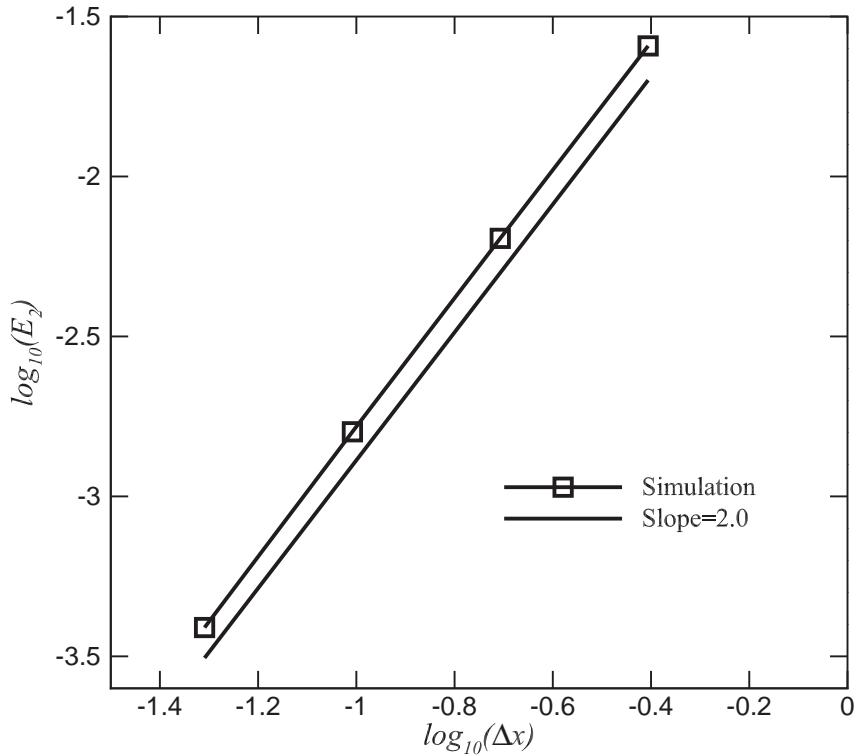,bbllx=20pt,bblly=20pt,bburx=540pt,bbury=480pt,width=0.7\textwidth,clip=}}	
	}
	\caption{Relative error $ {E_2} $ at different space step at Re=20.}
	\label{FIG2}
\end{figure*}

The relationship between grid size and $ {E_2} $ of the present scheme is presented in Fig. \ref{FIG2}, and the slope of the fitting line is 1.9953. The differences in $ {E_2} $ for different Reynolds numbers are very small (as shown in TABLE \ref{TAB1}), and all the slops are very close to 2. Besides, the slops do not change in different values for $ w_2 $, $ w_3 $, $ w_4 $ in our simulations. This demonstrates the scheme proposed has second-order accuracy in space.
\subsection{Thermal Couette flow}\label{sec5B}
To check the capability of describing viscous heat dissipation by the thermal cascaded LBM, two-dimensional thermal Couette flow is simulated. We consider the viscous fluid between two infinite parallel plates, in which the upper one is moving at speed $ U $ with temperature $ T_0 $ and the lower adiabatic plate is fixed. With the assumption that $ \Pr  = \mu {c_p}/\lambda $ is constant and $ (\mu /{\mu _0}) = (T/{T_0})
 $, there is an analytical solution \cite{Liepmann_1957},
\begin{subequations}
	\begin{equation}
(1{\rm{ + }}\Pr \frac{{\gamma  - 1}}{3}M{a^2})\frac{y}{H} = \frac{{{u_x}}}{U} + \Pr \frac{{\gamma  - 1}}{2}\text{Ma}^2\left[ {\frac{{{u_x}}}{U} - \frac{1}{3}{{(\frac{{{u_x}}}{U})}^3}} \right],
	\end{equation}
	\begin{equation}
	\frac{T}{{{T_0}}} = 1 + \Pr \frac{{\gamma  - 1}}{2}\text{Ma}^2\left[ {1 - {{(\frac{{{u_x}}}{U})}^2}} \right],
	\end{equation}
\end{subequations}
where $ \gamma  = (b + 2)/b
 $ is the specific-heat ratio, $ \text{Ma} = U/\sqrt {\gamma R{T_0}} 
  $ is the Mach number, and $ H $ is the distance between the two plates.
  
In our simulations, we set $ \text{Ma}=0.35 $ with different values of Pr and $ \gamma $: Pr=4 with $ \gamma  = 5/3(b = 3)
 $ and $ 3/2(b = 4)
  $; Pr=5 with $ \gamma  = 5/3(b = 3)
  $ and $ 3/2(b = 4)
  $. A uniform mesh $ {N_x} = {N_y} = 5 \times 40 $
  is employed. It should be noted that the mesh resolution is lower than that in \cite{Liluo_2012} but the same accuracy is achived (see results below). For the top and bottom walls, the non-equilibrium bounce-back method \cite{Zouhe} and non-equilibrium extrapolation method \cite{Guo_2007,Nonequ-extra} are adopted for velocity and temperature boundary conditions, respectively, while the periodic boundary condition is imposed along $ x $ direction. The upper wall temperature  and the reference temperature are set to unity, with a reference dynamic viscosity $ {\mu _0} = 0.35
   $. The first relaxation parameter $ w_1 $ is a field variable related to the local dynamic viscosity as given in Eq. (\ref{e29}), while others are constant with $ {w_2} = 1.1$, $ {w_3} = 1.6$, $ {w_4} = 0.3$.

\begin{figure*}[!ht]
	\center {
		{\epsfig{file=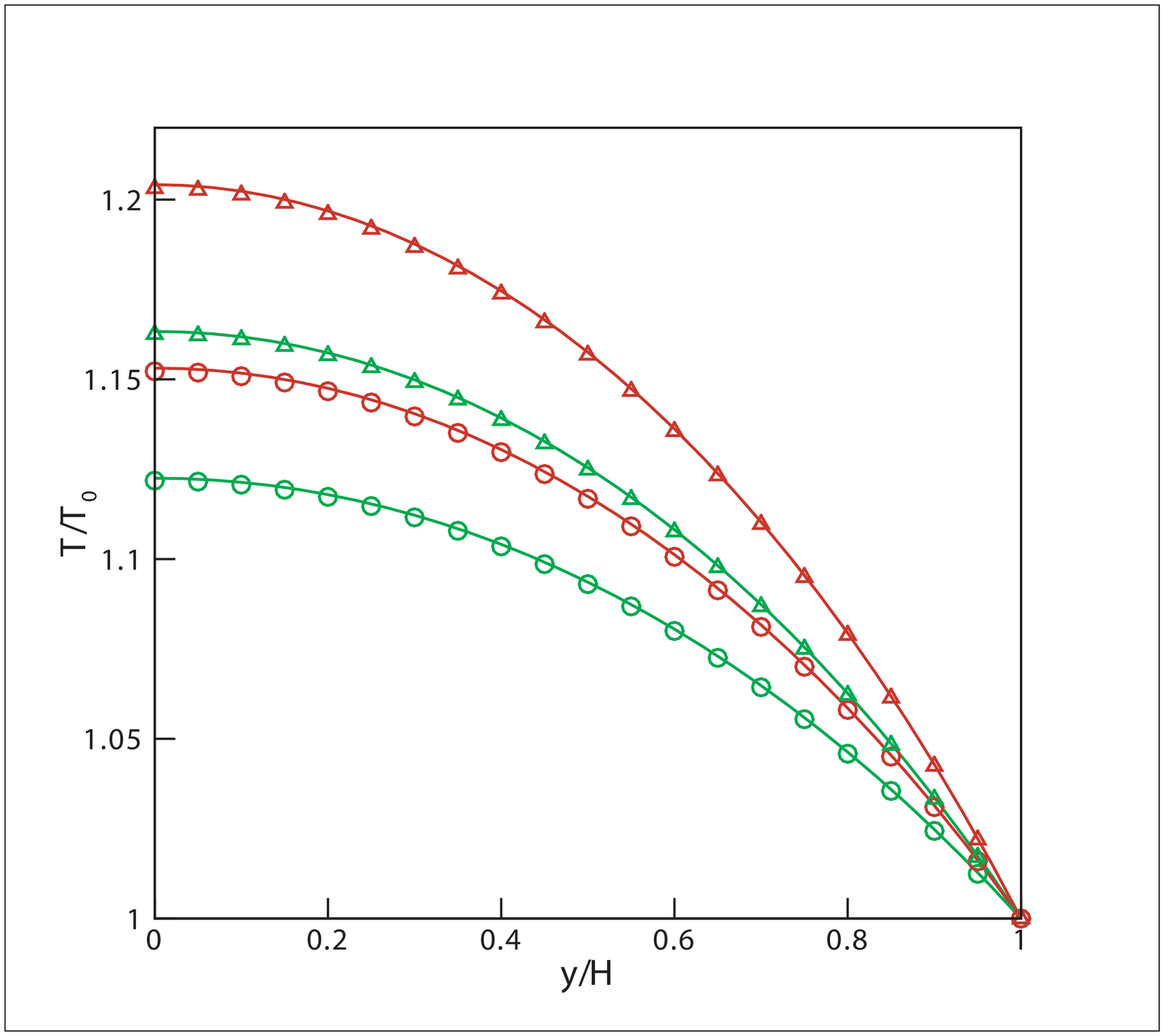,bbllx=25pt,bblly=25pt,bburx=545pt,bbury=485pt,width=0.5\textwidth,clip=}}	
	}
	\caption{Comparison between the simulation (symbols) and analytical (solid lines) results for dimensionless temperature profiles at $ \text{Ma}=0.35 $, with Pr=5 (red ones) and 4 (green ones). Deltas and circles are the simulation results for $ \gamma=5/3(b=3) $ and $ 3/2(b=4) $, respectively.}
	\label{FIG3}
\end{figure*}

\begin{table}[htbp]
	\caption{Comparison of the bottom wall temperatures between numerical and analytical results.}
	\label{TAB2}
	\begin{tabular}{ c   c   c   c   c}
		\toprule
		cases	&$ \text{Pr}=5, \gamma=5/3 $ & $ \text{Pr}=5, \gamma=3/2 $ & $ \text{Pr}=4, \gamma=5/3 $ & $ \text{Pr}=4, \gamma=3/2 $ \\ \hline
		$ T_n $	& 1.2031 &  1.1522  & 1.1626 & 1.1218 \\ \hline
		$ T_a $	& 1.2042  & 1.1531  & 1.1633 & 1.1225 \\ \hline
		$ E_r $	& 0.54 & 0.59  & 0.43 & 0.57 \\ 
		\toprule
	\end{tabular}
\end{table}    
Fig. \ref{FIG3} presents the simulation and analytical results for dimensionless temperature profiles in four cases. It can be observed that numerical results are in excellent agreement with the theoretical ones. To be specific, the temperatures at the bottom wall in numerical $ (T_n) $ and analytical $ (T_a) $
solutions are compared in TABLE \ref{TAB2}. The relative error is defined as $ E_r=(T_n -T_0)/(T_a - T_0) $. As presented in the TABLE \ref{TAB2}, the relative errors are less than $ 1\% $ in all the cases.
\subsection{Low-Mach Shock Tube Problem}
To check the present model's ability of simulating Low-Mach number compressible flow, a shock tube problem is studied in this section. The construction of this problem is that a long tube containing the same gas is separated by a barrier in the middle into two parts with different pressure, density and temperature. At the moment of removing the barrier, a complex flow is set up. The initial conditions for our simulations are,
\begin{equation}
\left\{ \begin{array}{l}
\left( {\rho /{\rho _0},{u_x}/{u_0},p/{p_0}} \right) = (1,0,0.2),~~~~0 \le x \le 0.5 \\ 
\left( {\rho /{\rho _0},{u_x}/{u_0},p/{p_0}} \right) = (0.5,0,0.1),~~0.5 < x \le 1 \\ 
\end{array} \right.
\end{equation}
where $ \rho _0=1.0 $, $ T_0=1.0 $, $ p_0=\rho_0 RT_0=1/3 $, $ {u_0} = \sqrt {{\rho _0}R{T_0}} $  are the reference density, temperature, pressure and velocity, respectively, and $ L_0 $ is the length of the tube.
\begin{figure*}[!ht]
	\center {
		{\epsfig{file=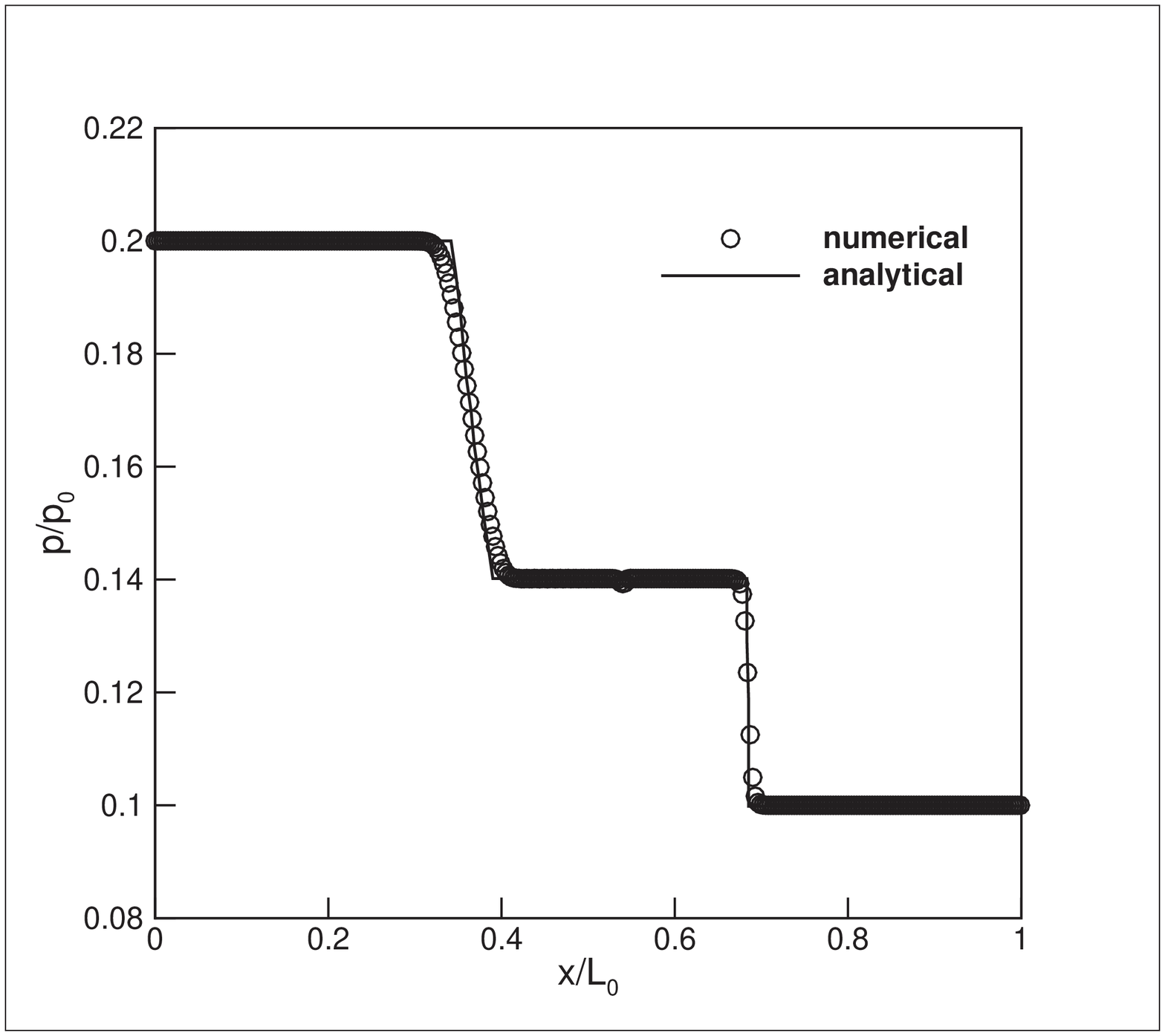,bbllx=20pt,bblly=20pt,bburx=540pt,bbury=490pt,width=0.4\textwidth,clip=}}\hspace{0.5cm}  
		{\epsfig{file=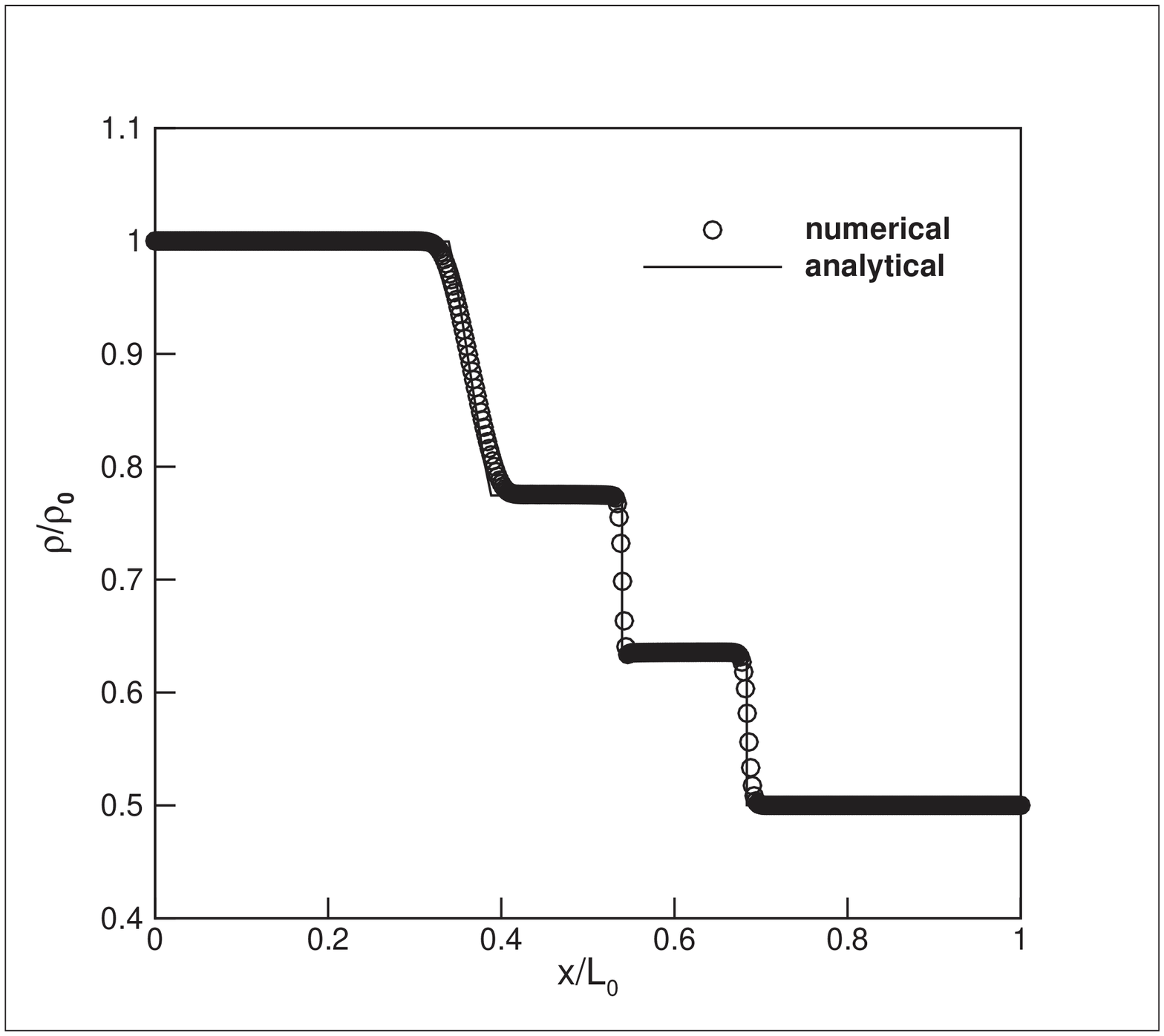,bbllx=20pt,bblly=20pt,bburx=540pt,bbury=490pt,width=0.4\textwidth,clip=}}\vspace{0.2cm}
		{\epsfig{file=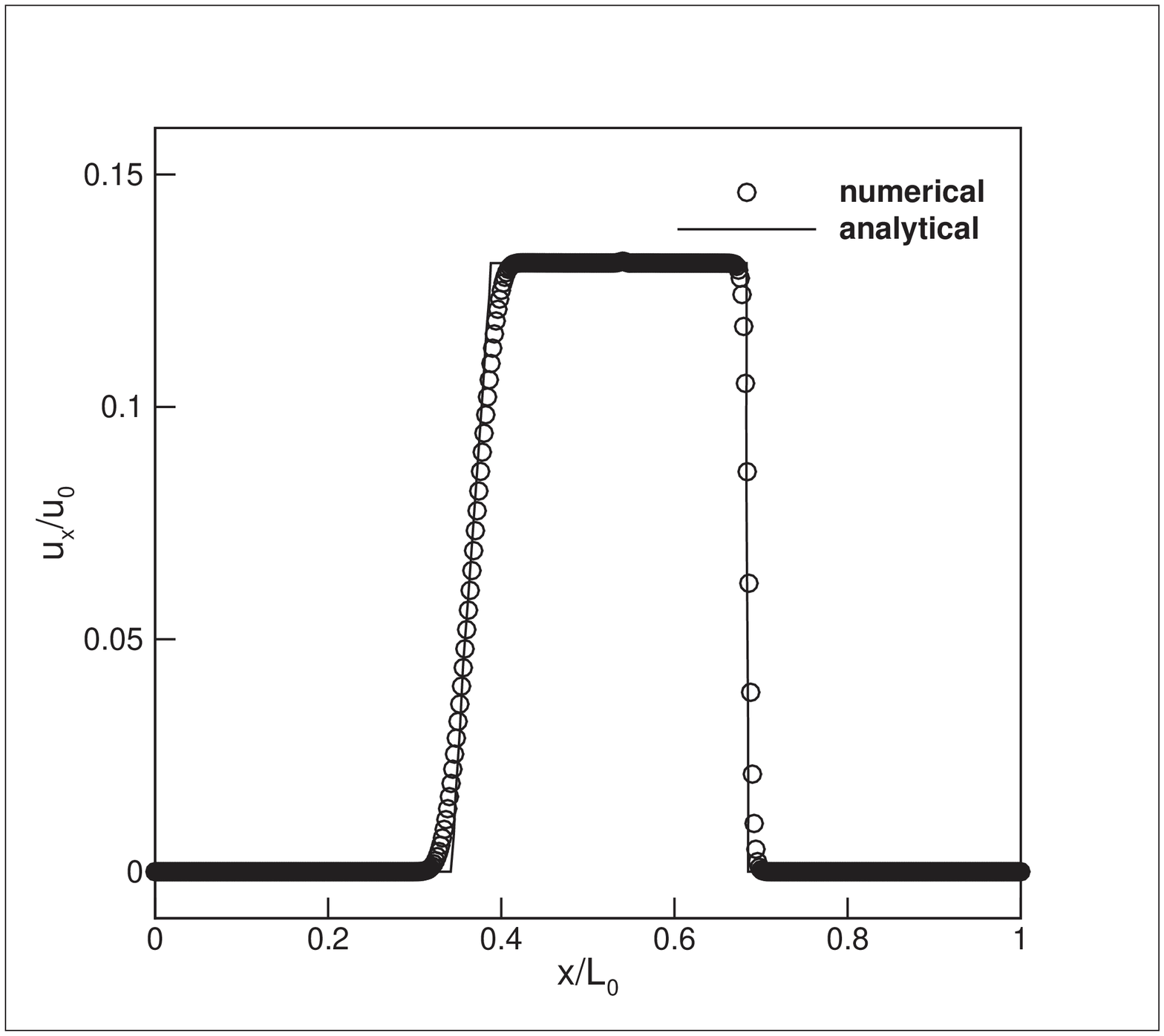,bbllx=15pt,bblly=20pt,bburx=540pt,bbury=490pt,width=0.4\textwidth,clip=}}\hspace{0.5cm}
		{\epsfig{file=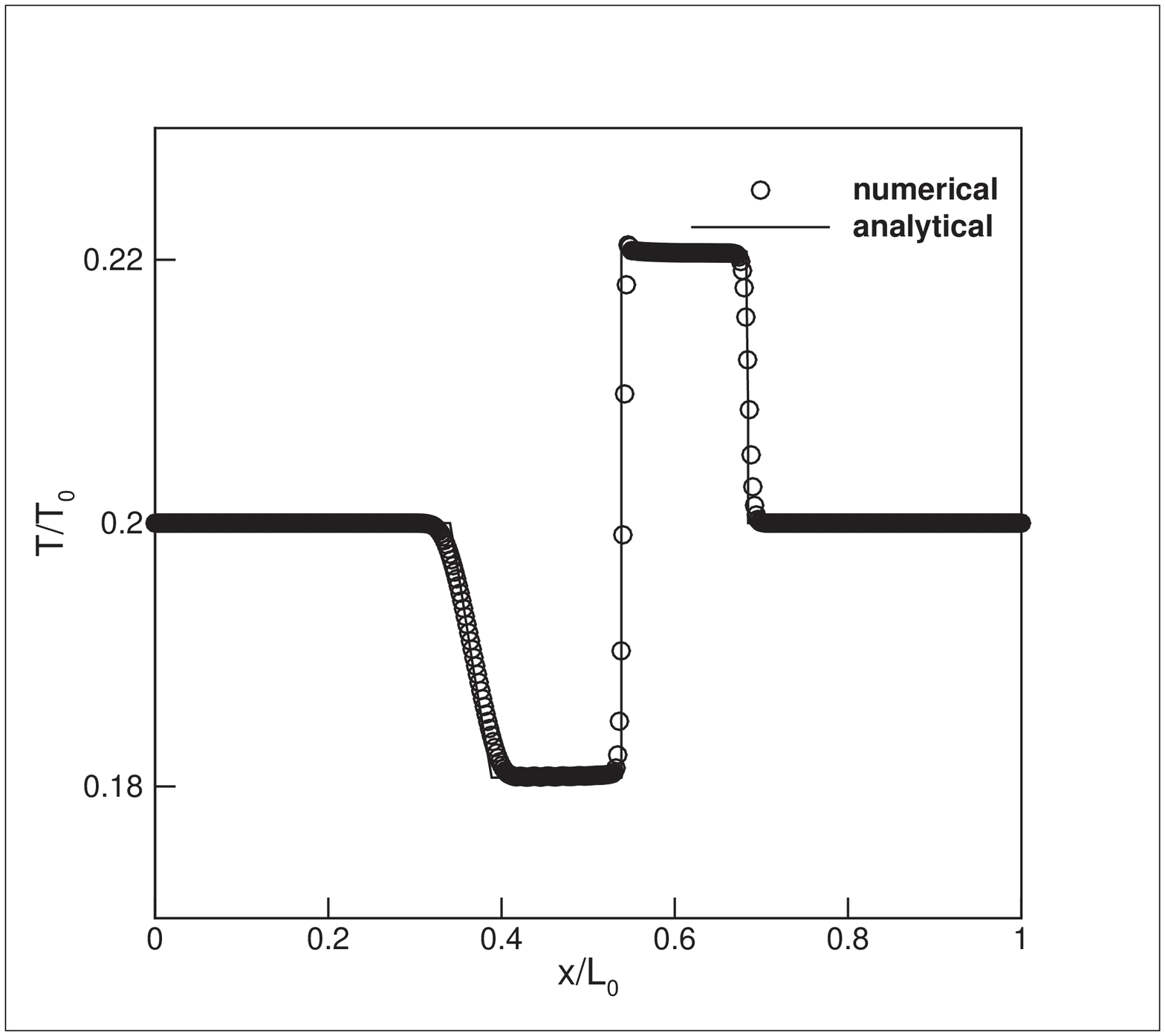,bbllx=15pt,bblly=20pt,bburx=540pt,bbury=490pt,width=0.4\textwidth,clip=}}
	}
\caption{Comparison between the simulation (symbols) and analytical \cite{Toro_1997} (solid lines) results for dimensionless density, pressure, horizontal velocity, and temperature profiles at $ t = 520\Delta t$. The symbol spacing is ISkip=2.}
\label{FIG04}
\end{figure*}

In the simulation, a $ 1000\times5 $ lattice is used, the periodic boundary condition is imposed along $ y $ direction, while EDFs are used in $ x=0 $ and $ x=1000 $. The specific heat ratio $ \gamma $ and Prandtl number Pr are set to 1.4 and 0.71, with $ w_1=1.887 $, and $ w_2=w_3=w_4=1.0 $. Simulation results are compared with the analytical ones in Fig. \ref{FIG04}. The four plots present dimensionless density, pressure, horizontal velocity, and temperature profiles, respectively, at time $ t = 520\Delta t$. It can be observed that numerical results are in good agreement with the theoretical ones.
\subsection{Rayleigh-B\'{e}nard convection}
To check the ability of simulating	thermal flow with external force by the present thermal model, the numerical experiment of the Rayleigh-B\'{e}nard convective flow is conducted in this section. The Rayleigh-B\'{e}nard instability is one of the classical thermal instability phenomena, in which the fluid is enclosed between two parallel stationary walls, the cold top and hot bottom, and experiences the gravity force. Linear stability theory has proven that convection develops most readily when the wave number is at the critical value 3.117 \cite{RB_Ref}, which is approximately corresponding to length-width ratio 2:1 in the flow domain.

Since the present model is a coupling model, we can implement the force by means of central moments with $ a_x=0 $ and $ a_y=-g $ ($ g $ is the gravity acceleration), without using the Boussinesq assumption. We conduct the experiment in the weakly compressible regime, with a 6$ \% $ temperature difference of the reference temperature $ T_0=1.0 $. To delete the heat dissipation term, we adopt the internal energy distribution function to simulate temperature field \cite{Liluo_2012,Liqing_2008}. The non-equilibrium extrapolation scheme \cite{Guo_2007,Nonequ-extra} is used to treat the upper and lower wall boundaries for both velocities and temperatures, whereas the periodic boundary scheme is applied along horizontal direction. The Prandtl number corresponds to air, Pr=0.71. Then the flow is characterized by the Rayleigh number Ra,
\begin{equation}
\text{Ra} = \frac{{g\beta ({T_l} - {T_h}){H^3}\Pr }}{{{\nu ^2}}},
\end{equation}
where $ T_l $ and $ T_h $ are the temperatures of upper and lower walls ($ T_l > T_h $), $ H  $ is the distance between the walls, and $ \nu 
 $ is the kinematic viscosity of the fluid. The characteristic velocity $ {u_c} = \sqrt {g\beta ({T_h} - {T_l})H} 
  $ in this flow  should be kept at an appropriate value, for example 0.08 in our simulation, to keep the flow in the low-Mach number regime. And $\beta $ is the thermal expansion coefficient, which is the reciprocal of reference temperature for the ideal gas considered here.
  
\begin{figure*}[!ht]
  	\center {
  		{\epsfig{file=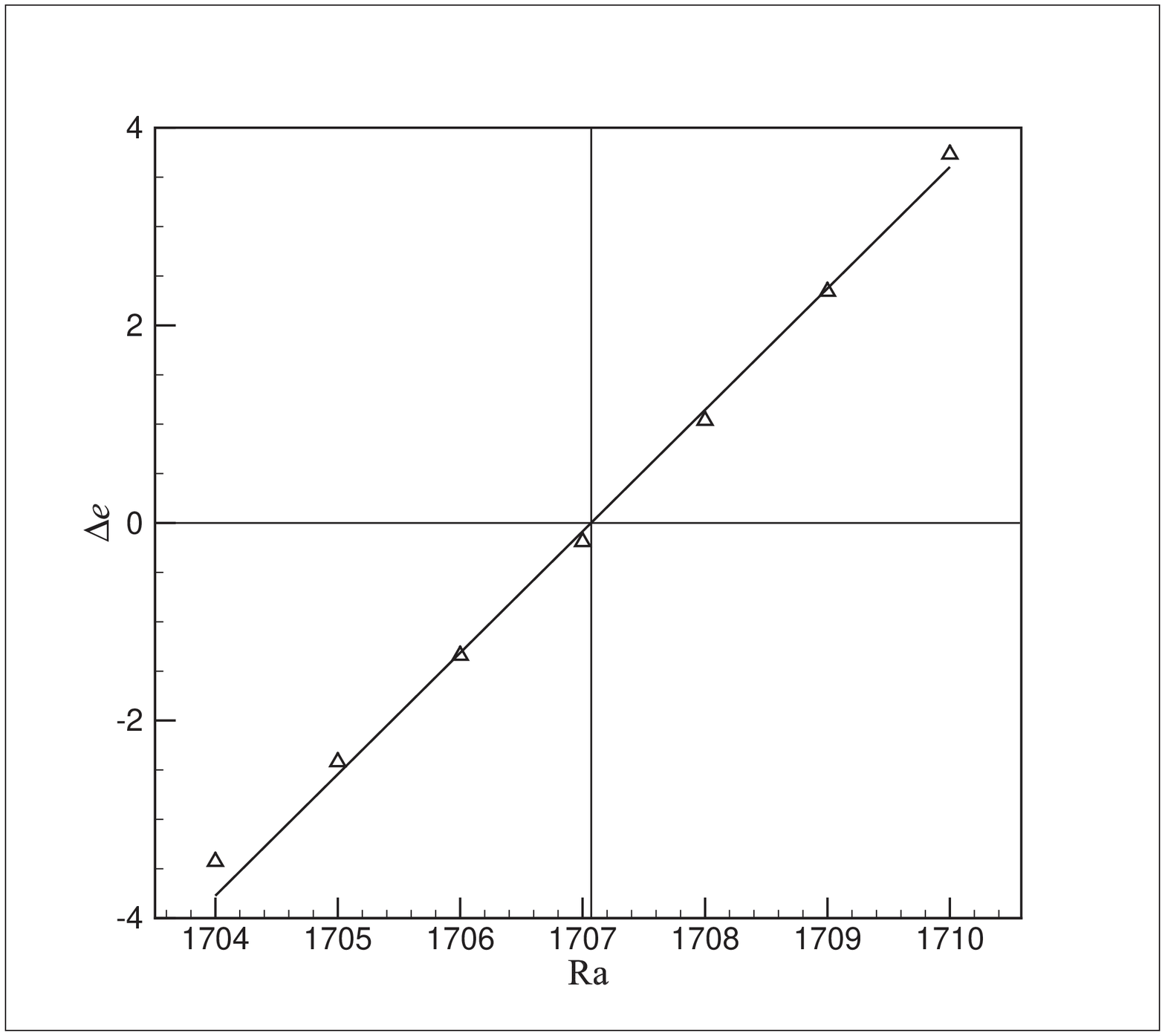,bbllx=40pt,bblly=25pt,bburx=540pt,bbury=480pt,width=0.5\textwidth,clip=}}\hspace{0.5cm}  
  	}
  \caption{Total internal energy increment $  \Delta e(\times10^{-10}) $ changes with different Rayleigh numbers. Triangles are the numerical results; the solid line is the linear fit for the simulations. The critical Rayleigh number extrapolated is $ \text{Ra}_\text{c}=1707.07 $.}
  \label{FIG05}
\end{figure*}

 For this kind of instability phenomenon, the driven force by the density variations induced by the temperature variations will balance with the viscosity force at the critical Rayleigh number $ \text{Ra}_\text{c} $, while if the Rayleigh number is increased above the threshold, the driving force will dominate and convection will start. First, we use a $ {N_x} \times {N_y} = 200 \times 100
  $ grid to calculate the critical Rayleigh number, and $ w_2=w_3=w_4=1.0 $ is used here and the following simulations. We initialize the temperature field with a linear distribution in y direction and give a small perturbation for density around the reference density $ T_0=1.0 $. It is noted that the total kinetic energy will keep increasing/decreasing lineally after the initial unsteady period around the critical Rayleigh number. For that, the total kinetic energy increment $ \Delta e$ every 10000 time steps in the domain is measured,
  \begin{equation}
  \Delta e = e(t + 10000) - e(t),~~e(t) = \sum\limits_{} {\left[ {\frac{1}{2}\rho (x,y,t){\mathbf{u}^2}(x,y,t)} \right]}.
  \end{equation}
  where $ \Delta e$ is measured by the slope of the total kinetic energy change with time, not at a certain time step. The critical Rayleigh number extrapolated is $ \text{Ra}_\text{c}=1707.07 $ (see Fig. \ref{FIG05}), which is in excellent agreement with the analytical value 1707.76.
\begin{figure*}[!ht]
	\center {
		{\epsfig{file=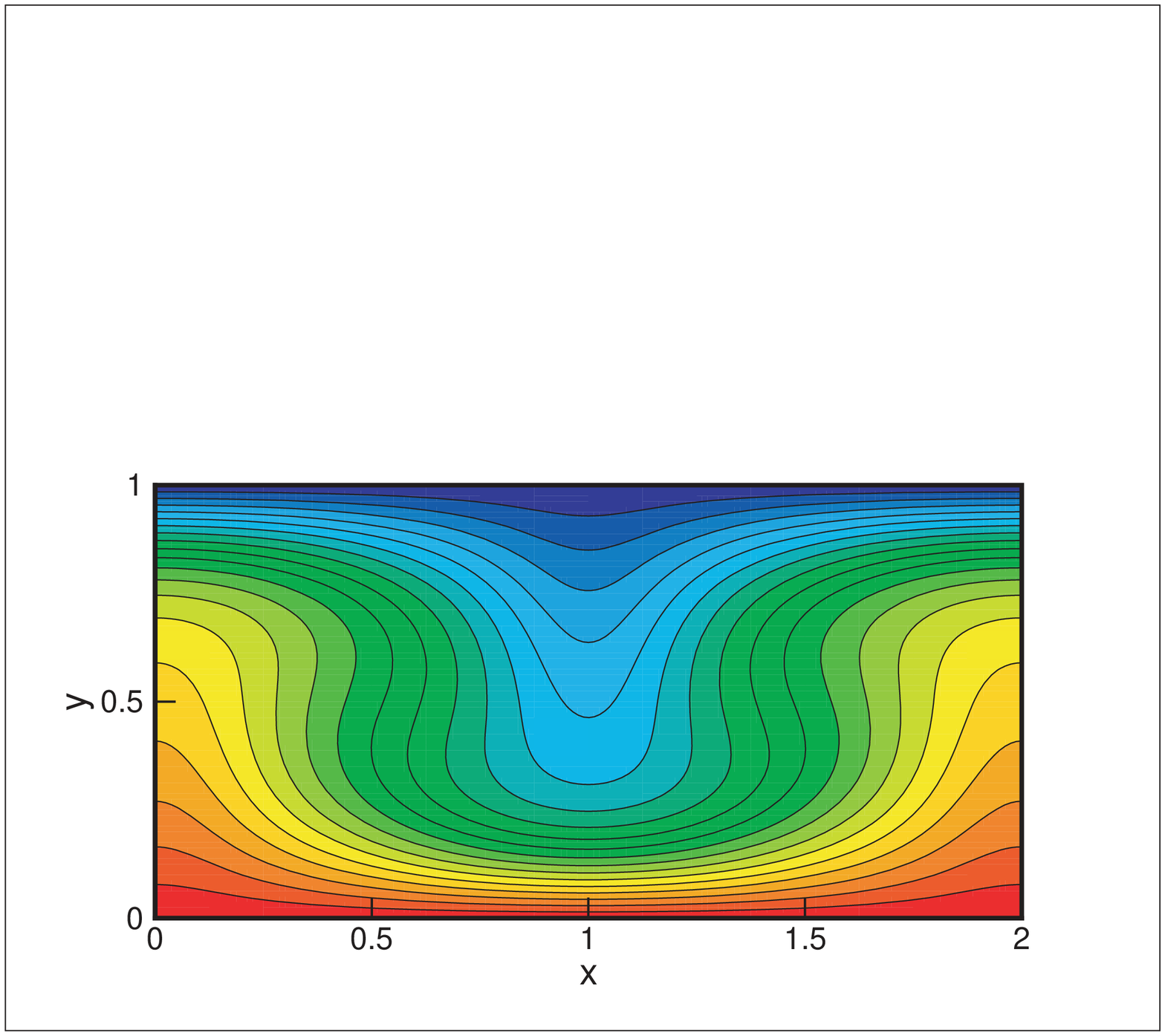,bbllx=35pt,bblly=25pt,bburx=540pt,bbury=300pt,width=0.5\textwidth,clip=}}\vspace{0.0cm}  
		{\epsfig{file=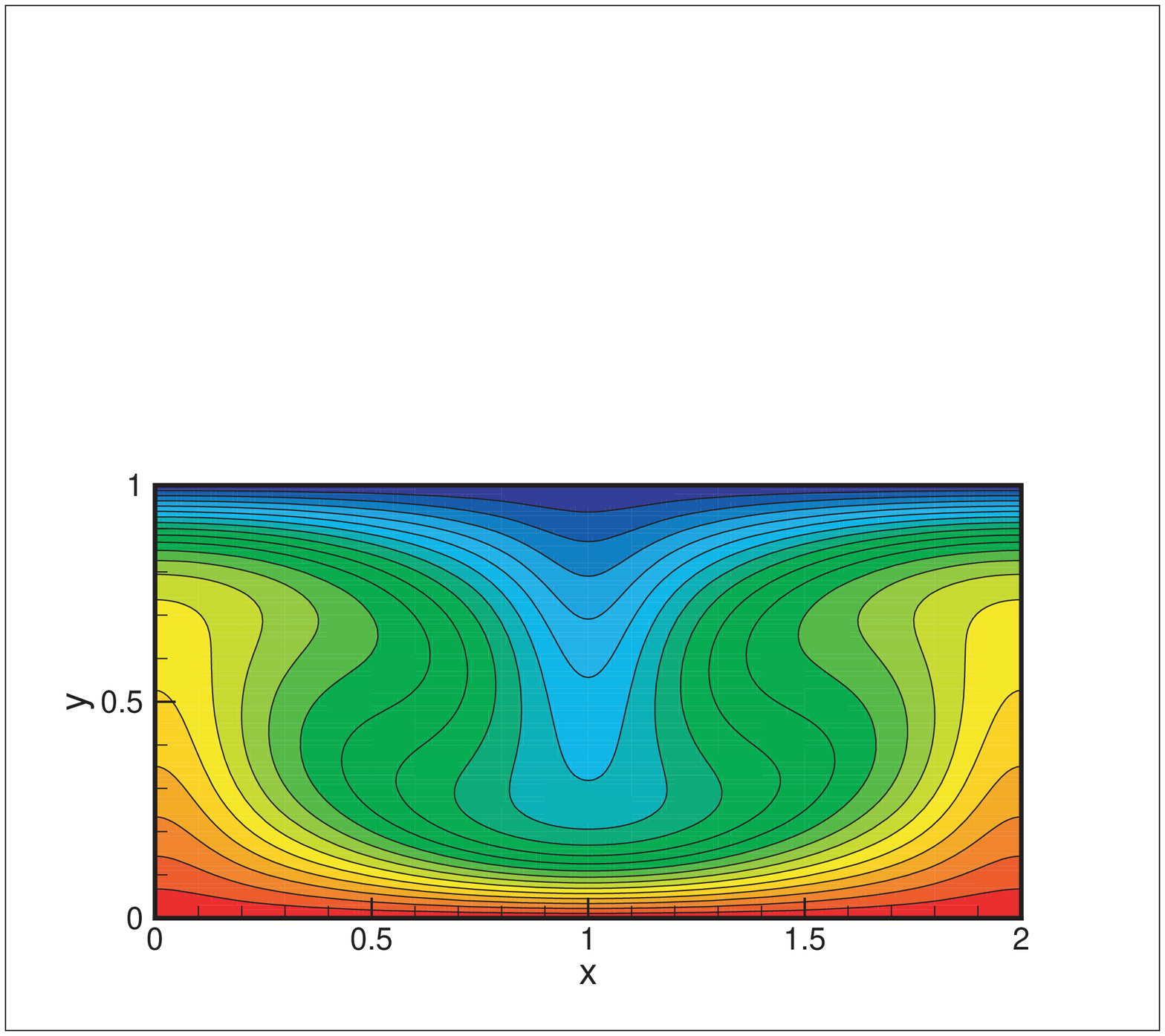,bbllx=37pt,bblly=25pt,bburx=540pt,bbury=300pt,width=0.5\textwidth,clip=}}\vspace{0.0cm} 
		{\epsfig{file=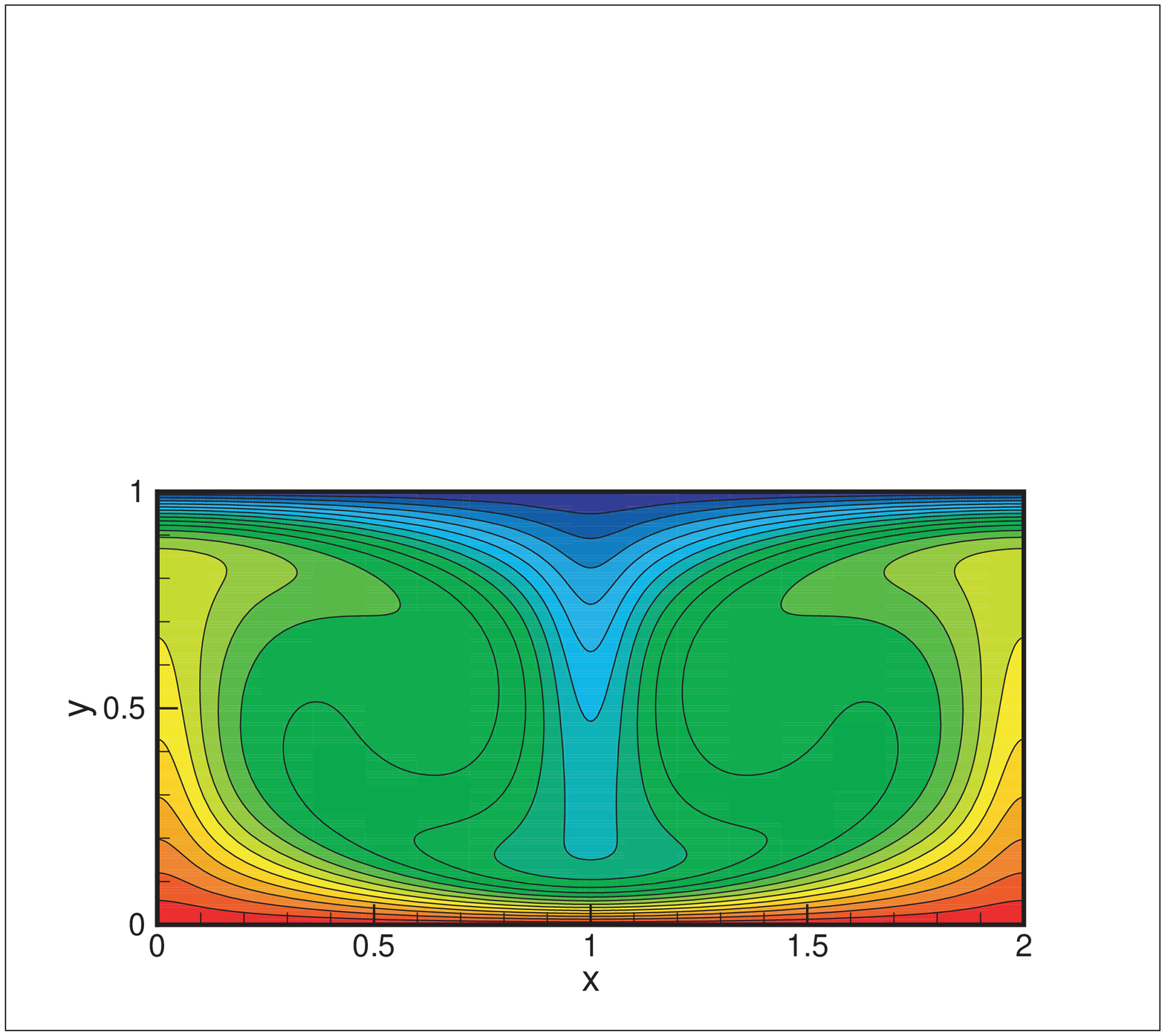,bbllx=35pt,bblly=25pt,bburx=540pt,bbury=300pt,width=0.5\textwidth,clip=}}
	}
	\caption{The normalized temperature $ (T-T_0)/\Delta T $ distribution for Rayleigh-B\'{e}nard convection flow. From top to bottom: Ra=5000, 10000 and 50000. A total of 19 equidistant lines are plotted.}
	\label{FIG06}
\end{figure*}

\begin{figure*}[!ht]
	\center {
		{\epsfig{file=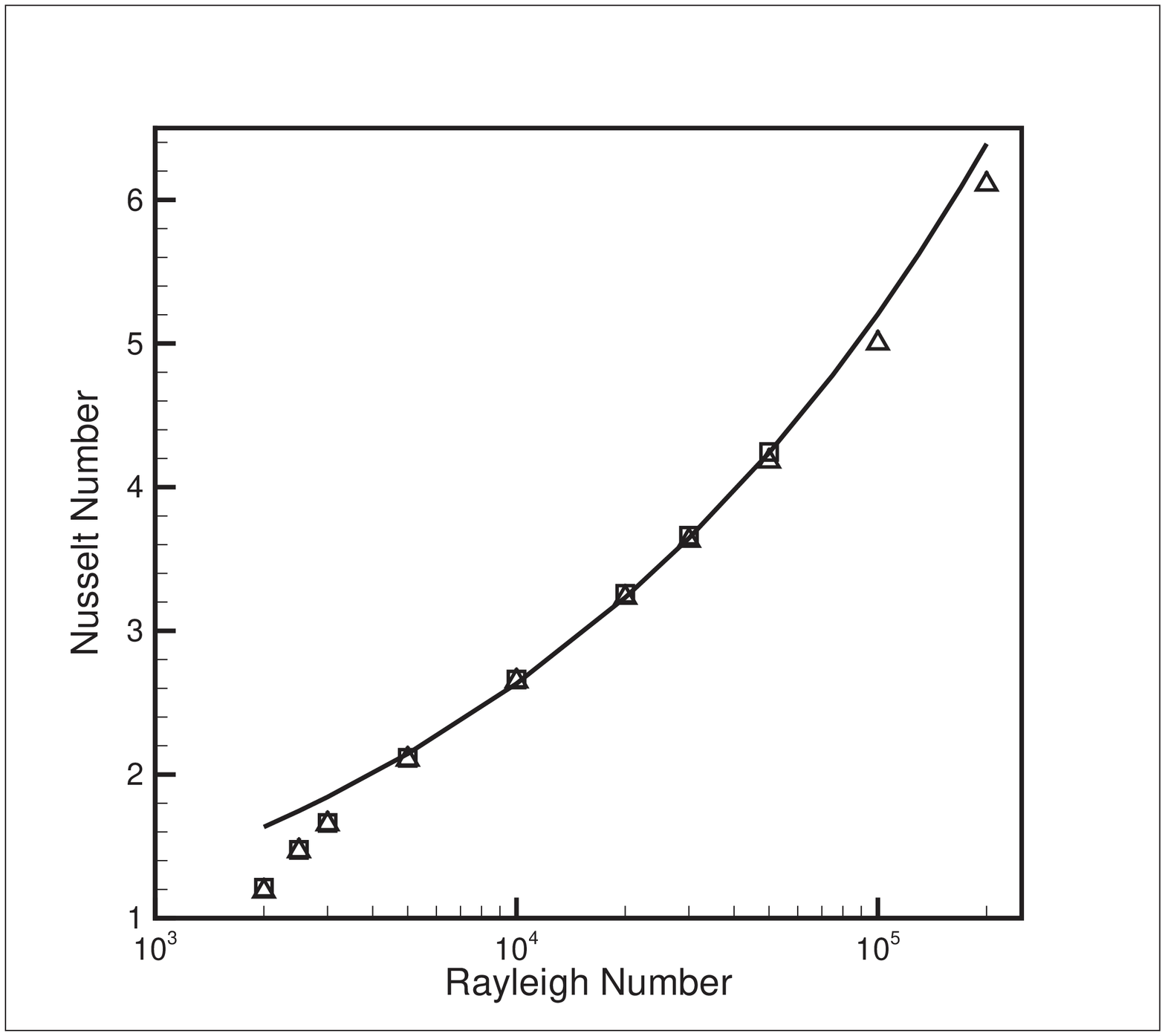,bbllx=30pt,bblly=15pt,bburx=550pt,bbury=480pt,width=0.5\textwidth,clip=}} 
	}
	\caption{Nusselt number vs Rayleigh number. Triangles: the present thermal CLBM; squares: reference data from Ref. \cite{RB_Ref}; and solid line: the empirical formula $ {\rm{Nu = 1}}{\rm{.56(Ra/R}}{{\rm{a}}_\text{c}}{{\rm{)}}^{0.256}}
		 $.}
	\label{FIG07}
\end{figure*}

Flows of different Rayleigh numbers are then simulated, $ N_y=80 $ is used if $ \text{Ra}<10000 $, $ N_y=100 $ if $ 10000\leq \text{Ra}<50000 $ and $  N_y=150 $ for $ \text{Ra}\geq 50000 $. The normalized temperature distribution for Rayleigh-B\'{e}nard convection at Ra=5000, 10000 and 50000 are shown in Fig. \ref{FIG06}. When the Rayleigh number increases, we can see two clear trends in the figures: the mixing of the hot and cold fluids is enhanced, and the temperature gradients near the bottom and top walls are increased, both of which mean the convective heat transfer is enhanced in the domain. To quantify this, the Nusselt number in the system is calculated,
\begin{equation}
\text{Nu} = 1 + \frac{{\left\langle {{u_y}(T - {T_0})} \right\rangle H}}{{k\Delta T}},
\end{equation}
where the square bracket represents the average over the whole system and $ k $ is the thermal conductivity of the fluid. The obtained values of Nusselt number $ \text{Nu}_\text{n}$ at various Rayleigh numbers are compared with the reference data in TABLE \ref{TAB3}, and plotted in Fig. \ref{FIG07}. The simulation results are in good agreement with those of Ref. \cite{RB_Ref} in a wide range of Ra as given in TABLE \ref{TAB3}. During the small Rayleigh number range (Ra<5000), convention is suppressed so that the Nusselt number decreases rapidly to 1.0 at $ \text{Ra}=\text{Ra}_\text{c} $, where the empirical formula loses efficacy. At very high Rayleigh numbers, the numerical results slightly underestimate the heat transfer, while this trend was also observed in other LBM studies \cite{He_1998,Karlin_2007,Feng_2015}.

\begin{table}[htbp]
	\caption{Comparison of Nusselt number between the present numerical results and the results in Ref. \cite{RB_Ref}.}
	\label{TAB3}
	\begin{tabular}{ c   c   c   c   c   c}
		\toprule
		cases	& $ \text{Ra}=2500 $ & $ \text{Ra}=5000 $ & $ \text{Ra}=10000 $ & $ \text{Ra}=30000 $ & $ \text{Ra}=50000 $\\ \hline
		$ \text{Nu}_n$ & 1.468 & 2.106 & 2.650 & 3.629 & 4.181 \\ \hline
		 Ref. \cite{RB_Ref}	&1.475  & 2.116 &2.661 & 3.662 & 4.245 \\ \hline
		$ {E_R}(\% )$	& 0.47 & 0.47  & 0.41 & 0.90 &1.51 \\ 
		\toprule
	\end{tabular}
\end{table}  

\section{CONCLUSIONS}\label{sec.6}
In this paper, we developed a method of incorporating force term into CLBM, and extended CLBM to thermal flow in the framework of the DDF approach.

To introduce the effect of external force, we match the discrete central moments of the force source term with the continuous central moments of the change of continuous DF by the force field, and the same transformed DF as done in \cite{Premnath_2009} was used to remove the implicitness, thus the present force scheme is consistent with that one. Numerical simulations have been conducted in a flow (steady Taylor-Green flow) with spatially varying body force and the second-order accuracy in space for the force scheme is verified. Compared with standard LBGK with the force method of Guo \cite{Guoforce}, the performance of CLBM with the present force method is better due to inclusion of higher order velocity terms. Though the force scheme of Guo also shows very good accuracy by using the EDF of CLBM, it is limited to the BGK relaxation ($ {w_1} = {w_2} = {w_3} = {w_4} $). Thus the present scheme is preferred to match CLBM with independent relaxation rates for different central moments.

Considering the respectable numerical performances for DDF-based thermal LBMs, we constructed the thermal CLBM in this framework. Firstly, the reference temperature in equilibrium central moments for the density DF was replaced by the local temperature. Secondly, a correction term was introduced similarly by means of central moments to remove the derivation of two diagonal elements for the third-order raw moments. Then a total energy EDF was introduced according to the required raw moments. Finally, by relaxing the density DF and total energy DF using the cascaded and BGK schemes respectively, a DDF thermal CLBM was constructed, where the density DF solves flow field and total energy DF solves temperature field and they are coupled naturally by EOS for the ideal gas. For thermal flows, a thermal Couette flow was simulated first, and the simulation results agreed well with the analytical solutions in different cases, which demonstrated the present thermal CLBM can include viscous heat dissipation with adjustable Prandtl number and specific-heat ratio. The TCLBM's ability of researching low-Mach compressible flows was then verified by simulating a low-Mach shock tube problem. Moreover, the numerical results for the Rayleigh -B\'{e}nard convection confirmed that the model can simulate thermal problems with force field without invoking the Boussinesq assumption.

In summary, unlike many multispeed-based coupling thermal LB models, the present model adopts the DDF approach on the standard lattice and thus retains  good features of LBM, such as simplicity and numerical efficiency. Moreover, different from many previous DDF LB models where passive scalar model and/or Boussinesq assumption are used to simulate the temperature field, the present model is a coupling model and can be used for more general thermal flows. Finally, the present TCLBM can be extented to three dimensions (3D) readily based on the 3D cascaded LBM \cite{Luokaihong_2014b}.

\section*{Acknowledgments}

Linlin Fei would like to gratefully acknowledge helpful discussions with Dr. Qing Li and Dr. Chuandong Lin. Support from the National Key Research and Development Plan (No. 2016YFB0600805) and the Center for Combustion Energy at Tsinghua University is gratefully acknowledged.

\end{document}